\documentclass{jfm}
\usepackage{natbib}

\let\realverbatim=\verbatim
\let\realendverbatim=\endverbatim
\renewcommand\verbatim{\par\addvspace{6pt plus 2pt minus 1pt}\realverbatim}
\renewcommand\endverbatim{\realendverbatim\addvspace{6pt plus 2pt minus 1pt}}
\makeatletter
\newcommand\verbsize{\@setfontsize\verbsize{10}\@xiipt}
\renewcommand\verbatim@font{\verbsize\normalfont\ttfamily}
\makeatother

\ifCUPmtlplainloaded \else
  \checkfont{eurm10}
  \iffontfound
    \IfFileExists{upmath.sty}
      {\typeout{^^JFound AMS Euler Roman fonts on the system,
                   using the 'upmath' package.^^J}%
       \usepackage{upmath}}
      {\typeout{^^JFound AMS Euler Roman fonts on the system, but you
                   don't seem to have the}%
       \typeout{'upmath' package installed. jfm.cls can take advantage
                 of these fonts,^^Jif you use 'upmath' package.^^J}%
       \providecommand\mathup[1]{##1}%
       \providecommand\mathbup[1]{##1}%
      }
  \else
    \providecommand\mathup[1]{#1}%
    \providecommand\mathbup[1]{#1}%
  \fi
\fi

\ifCUPmtlplainloaded \else
  \checkfont{msam10}
  \iffontfound
    \IfFileExists{amssymb.sty}
      {\typeout{^^JFound AMS Symbol fonts on the system, using the
                'amssymb' package.^^J}%
       \usepackage{amssymb}%
       \let\le=\leqslant  
       \let\ge=\geqslant  
      }{}
  \fi
\fi

\ifCUPmtlplainloaded \else
  \IfFileExists{amsbsy.sty}
    {\typeout{^^JFound the 'amsbsy' package on the system, using it.^^J}%
     \usepackage{amsbsy}}
    {\providecommand\boldsymbol[1]{\mbox{\boldmath $##1$}}}
\fi

\newsavebox{\astrutbox}
\sbox{\astrutbox}{\rule[-5pt]{0pt}{20pt}}

\title[Journal of Fluid Mechanics]{Multifractal nature of plume structure in high Rayleigh number convection}

\author[B. A. Puthenveettil, G. Ananthakrishna and J. H. Arakeri ]{B\ls A\ls B\ls U\ls R\ls A\ls J\ns A.\ns P\ls U\ls T\ls H\ls E\ls N\ls V\ls E\ls E\ls T\ls T\ls I\ls L,\ns$^1$ \\G.\ns A\ls N\ls A\ls N\ls T\ls H\ls A\ls K\ls R\ls I\ls S\ls H\ls N\ls A\ns$^2$\and J\ls A\ls Y\ls W\ls A\ls N\ls T\ns H.\ns A\ls R\ls A\ls K\ls E\ls R\ls I\ns$^1$}

\affiliation{$^1$Department of Mechanical Engineering, $^2$Materials Research Center, and Centre for Condensed Matter Theory (garani@mrc.iisc.ernet.in),\\  Indian Institute of Science, Bangalore, India.}
\date{July 2004}
\pubyear{2001} 
\volume{000}
\pagerange{\pageref{firstpage}--\pageref{lastpage}}
\doi{S002211200100456X}
\usepackage{graphicx} 
\usepackage{subfigure} 
\usepackage{verbatim} 
\usepackage{nomencl} 
\begin{document}
\bibliographystyle{jfm}
\label{firstpage}
\maketitle
\begin{abstract}
  The geometrically different plan forms of near wall plume structure
  in turbulent natural convection, visualised by driving the
  convection using concentration differences across a membrane, are
  shown to have a common multifractal spectrum of singularities for
  Rayleigh numbers in the range $10^{10}- 10^{11}$ at Schmidt number
  of 602. The scaling is seen for a length scale range of $2^5$ and is
  independent of the Rayleigh number, the flux, the strength and
  nature of the large scale flow, and the aspect ratio.  Similar
  scaling is observed for the plume structures obtained in the
  presence of a weak flow across the membrane. This common non trivial
  spatial scaling is proposed to be due to the same underlying
  generating process of the near wall plume structures.
\end{abstract}
\section{Introduction}
\label{sec:introduction}
Turbulent Rayleigh - B\'{e}nard convection and other related natural
convection flows at high Rayleigh numbers have many unresolved issues;
these are extensively discussed in the review of \cite{siggr}.  As
most high $Ra$ studies are conducted in cryogenic conditions
\cite*[][]{niemn}, which prevent visualisation of the near wall
structures, little is known about the nature of near wall coherent
structures in high Rayleigh number turbulent free convection. The
non-dimensional parameters that characterise natural convection flows
are the Rayleigh number ($Ra =g\frac{\Delta\rho}{\rho}H^3/\nu \kappa$,
the ratio of buoyancy effects to dissipative effects); Prandtl number
($Pr=\nu/\kappa$ , a fluid property); Nusselt number (
$Nu=\frac{Q}{\kappa \Delta T/H}$, a non-dimensional flux) and aspect
ratio (AR = $L/H$ , a geometric parameter of the fluid layer) where,
$g$ = the acceleration due to gravity, $\frac{\Delta\rho}{\rho}$ =
nondimensional density difference, $Q$ = the kinematic heat flux,
$\Delta T$ = the temperature difference between the walls, $H$ = fluid
layer height, $L$ = fluid layer width, $\nu$= kinematic viscosity and
$\kappa$ = thermal diffusivity.

Most of the previous studies on near wall coherent structures give
only a qualitative picture of closed polygonal forms of line plumes at
$Ra \sim 10^6$, that change to randomly moving and merging line plume
patterns at higher $Ra$ of $10^7 -10^8$
\cite*[][]{sprow,tas,tapf,kerr2}. Line plumes are buoyant fluid rising
in the form of sheets from lines on the horizontal heated surface.
Visualisations at the highest $Ra$ so far ($\sim 10^{9}$) by
\cite*{zocchi} show line plumes being swept along the direction of
shear caused by the large scale flow.  \cite{tjfm} showed that the
randomly spaced and oriented plumes near the wall can be modeled as a
regular array of laminar line plumes, each plume fed by laminar
natural convection boundary layers on either side. \cite{mine}
extended the analysis to high $Ra$- high $Pr$ case, and found that the
probability distribution of plume spacings have a common log-normal
form, independent of the parameter values. None of these studies
address the spatial scaling of the plume structure pattern.

In this paper, we report the spatial scaling of the near wall
plan-form plume structures in high Rayleigh number turbulent
free-convection, using multifractal formalism. Briefly, multifractal
formalism describes the statistical properties of singular measures in
terms of the singularity spectrum $f(\alpha)$, corresponding to the
singularity strength $\alpha$. $f(\alpha)$ can be regarded as the
fractal dimensions of subsets with corresponding singularity strength
$\alpha$. The details of multifractal formalism can be seen in
~\cite{mb2,ms}; \cite*{hals}, and the references cited therein.  Even
though studies abound on applying this formalism for characterising
spatial structures in various fields, we are not aware of such an
analysis of near wall coherent structures of turbulent convection.

\section{Experimental setup and image preprocessing}
\label{sec:exp}
\begin{figure}
\centering
\includegraphics[width=0.6\textwidth]{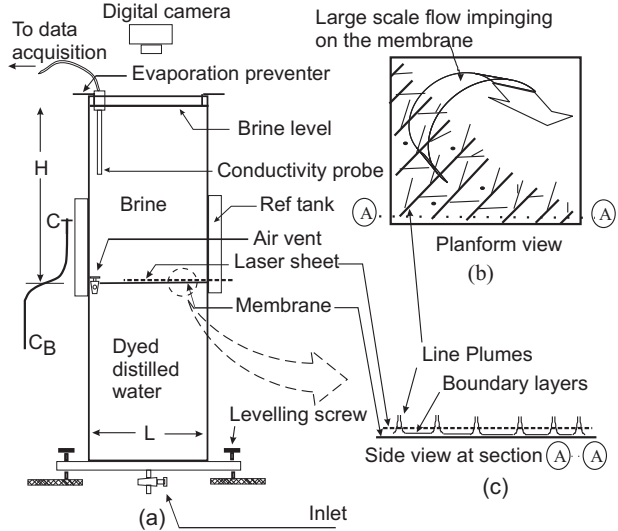}  \label{fig:setup}
  \caption{(a) The experimental  set up (b) The plan form view of the near wall plume structure as seen at the intersection of a horizontal laser sheet. (c) The side view in a vertical plane showing rising line plumes. Planform in (b) is for the TF-type convection showing a plume free area where the large scale flow impinges on the membrane.}
\end{figure}
We visualise the near wall plume structure at high $Ra$ by driving the
convection using concentration differences of NaCl across a membrane.
Concentration in the present experiments is equivalent to temperature
in Rayleigh - B\'{e}nard convection. High $Ra$ ($10^{10} -10 ^{11}$)
are achieved at Schmidt number ($Sc$ $\sim \nu/D$, equivalent to $Pr$)
of 602 due to the low molecular diffusivity of NaCl. The set-up
consists of two glass compartments of square cross section, arranged
one on top of the other with a fine membrane fixed horizontally in
between them; the schematic is shown in Figure \ref{fig:setup}. The
membranes used are Pall Gelmann\texttrademark NX29325 membrane disc
filters (PG henceforth) with a random pore structure having a mean
pore size of $0.45\mu$ and Swedish Nylobolt 140S screen printing
membrane (140s henceforth) with a regular square pore of size $35\mu
\times 35\mu$.

The bottom tank is filled with distilled water tagged with a small
amount of Sodium Fluoresceine (absorption spectra peak at 488nm and
emission spectra peak at 518nm) and then the top tank filled with
brine to initiate the experiment.  The convection is unsteady, but
quasi-steady approximation can be used as the time scale of change of
the density difference $(\frac{\Delta\rho}{\rho} /\frac{d
  \frac{\Delta\rho}{\rho}}{dt})$ is 10 times the time scale of one
large scale circulation $(H/W_*)$, where $W_* \sim \left(g\beta Q
  H\right)^{1/3}$\cite*[][]{dear1} is an estimate of the large scale
flow strength, H is the top tank fluid layer height, $Q$ denotes the
flux of NaCl and $\beta$ is the coefficient of salinity.  A horizontal
laser sheet, expanded and collimated from a 5W Spectra Physics
Stabilite\texttrademark 2017 Ar- Ion laser at 488nm is passed just
above ($<$1 mm) the membrane.  The dye in the bottom solution while
convecting upward fluoresces on incidence of the laser beam to make
the plume structure visible. A schematic of the visualisation of the
near wall phenomena is shown in Figure \ref{fig:setup}(b) and (c).  A
visible long pass filter glass, Coherent optics OG-515, is used to
block any scattered laser light and allow the emitted fluorescence to
pass through.  The images are captured on a digital handycam Sony DCR
PC9E. Experiments are conducted in 23 cm high tanks, with one tank
having 15 cm$\times$15 cm (AR = 0.65) cross section and another with
10 cm$\times$10 cm (AR = 0.435) cross section.  Starting top tank NaCl
concentrations of 10g/l, 7g/l and 3g/l are used to study the plume
structure under different $Ra$.

The Laser Induced Fluorescence (LIF) images are RGB 24bit color with
640 $\times$ 480 pixels resolution.  The multifractal analysis is
conducted on binary images obtained from these RGB images.  In using
binary images, we are neglecting the intensity variation of the
fluorescence (proportional to the concentration of the dye) within the
plume line thickness.  The analysis is hence valid only for the
geometrical aspects of the planform.  A similar telegraph
approximation of the temperature time trace measured at the middle of
the fluid layer has been studied by \cite*{berd} to show the
clusterisation of plumes. The RGB image is cropped to remove the tank
walls, converted to gray scale, Radon transformed to remove the lines
(formed due to imperfections in the optics and the test section
surface) and then re-sampled to increase the resolution.  The
non-uniform background illumination due to the attenuation of the
laser sheet is subtracted to obtain the plume lines over a uniform
dark background.  Contrast and intensity are enhanced to make the
plume lines clearer before converting to a black and white binary
image using a threshold.  The effect of thresholding on the
multifractal exponents is considered in
Appendix~\ref{sec:effect-thresholding}. The binary image retains
almost all the features of the raw image.
\section{The plume structure planforms}
\label{sec:plume-struct-planf}
\begin{figure}
  \centering \subfigure[PG membrane in 15 cm$\times$15 cm tank, $Ra$ =
  2.04$\times 10^{11}$, $\Delta C$ = 3.2 g/l, $Q$ = 0.1 mg/cm$^2$/min,
  $W_*$ = 0.31 cm/s, $Re$=797, Image area = 14.65cm square, 714$^2$pix
  ]{\includegraphics[width=0.3\textwidth]{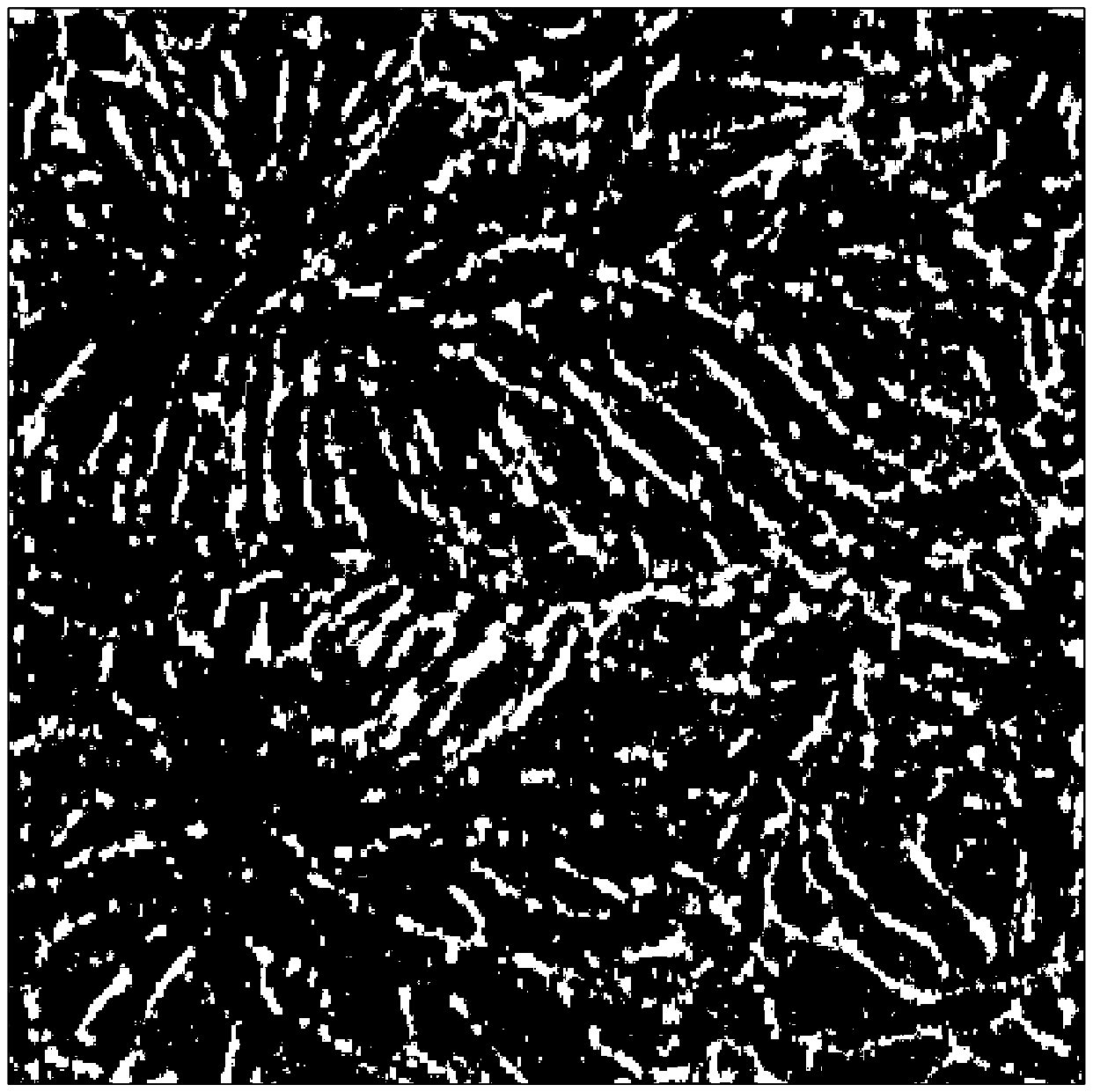}\label{fig:10j1}}
  \hfill \subfigure[Structure after 30seconds from (a) $Ra$ = 2.03$\times
  10^{11}$, $\Delta C$ = 3.19 g/l, $Q$ = 0.1 mg/cm$^2$/min, $W_*$ =
  0.31 cm/s, $Re$ = 797, Image area = 13.74 cm square, 728$^2$ pix
  ]{\includegraphics[width=0.3\textwidth]{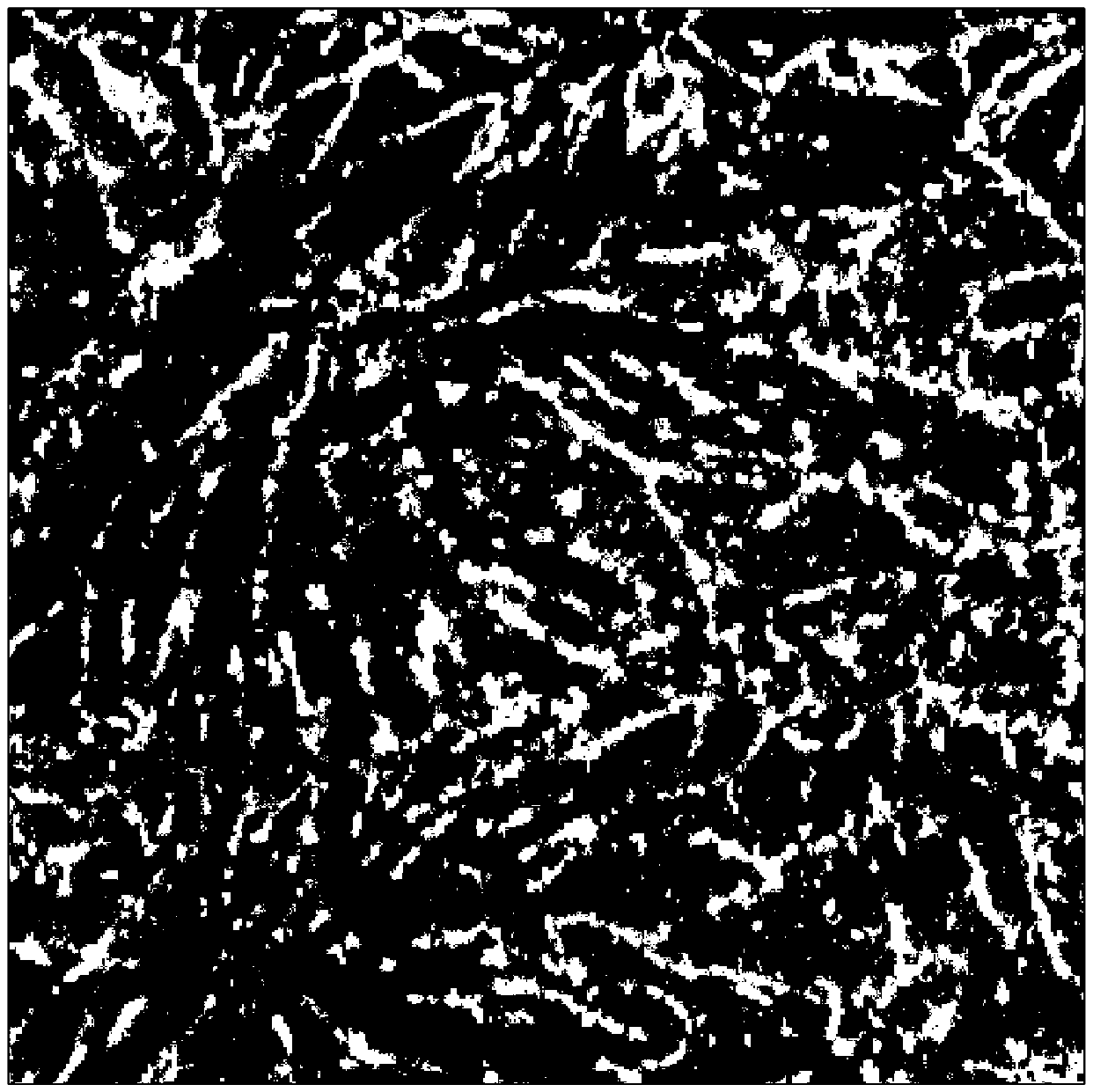}\label{fig:10j2}}
  \hfill \subfigure[PG membrane in 10 cm$\times$10 cm tank,
  $Ra$ = 2.03$\times 10^{11}$, $\Delta C$ = 3.198 g/l, $Q$ = 0.11
  mg/cm$^2$/min, $W_*$ = 0.31cm/s, $Re$ = 797, Image area = 9.157 cm
  square, 728$^2$ pix]{\includegraphics[width=0.3\textwidth]{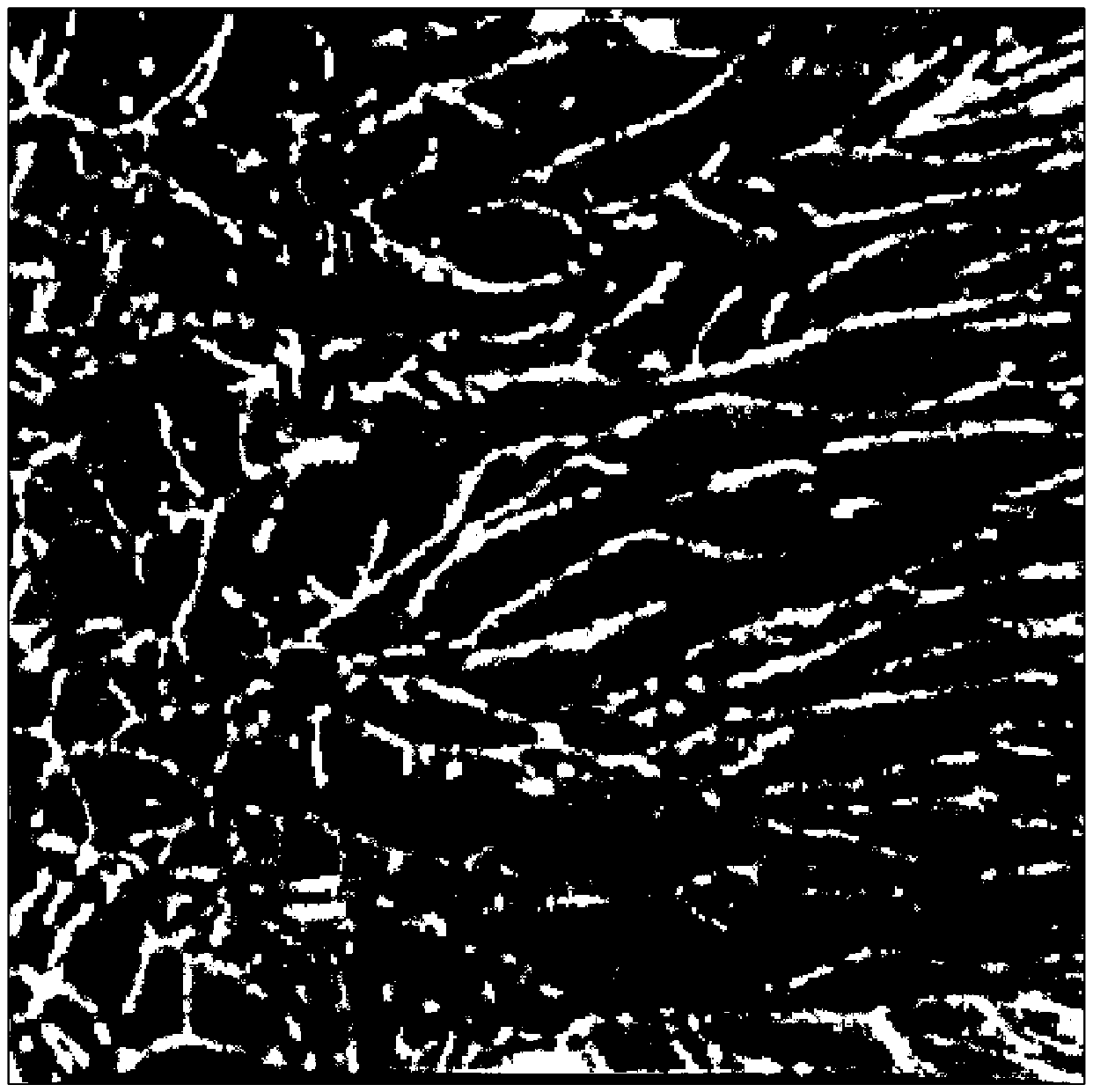}\label{fig:18d}}
  \hfill \subfigure[PG membrane in 15 cm$\times$15 cm tank,
  $Ra$ = 5.56$\times 10^{10}$, $\Delta C$ = 0.875 g/l, $Q$ = 0.02
  mg/cm$^2$/min, $W_*$ = 0.18cm/s, $Re$ = 463.6, Image area = 12.83cm
  square,
  728$^2$ pix]{\includegraphics[width=0.3\textwidth]{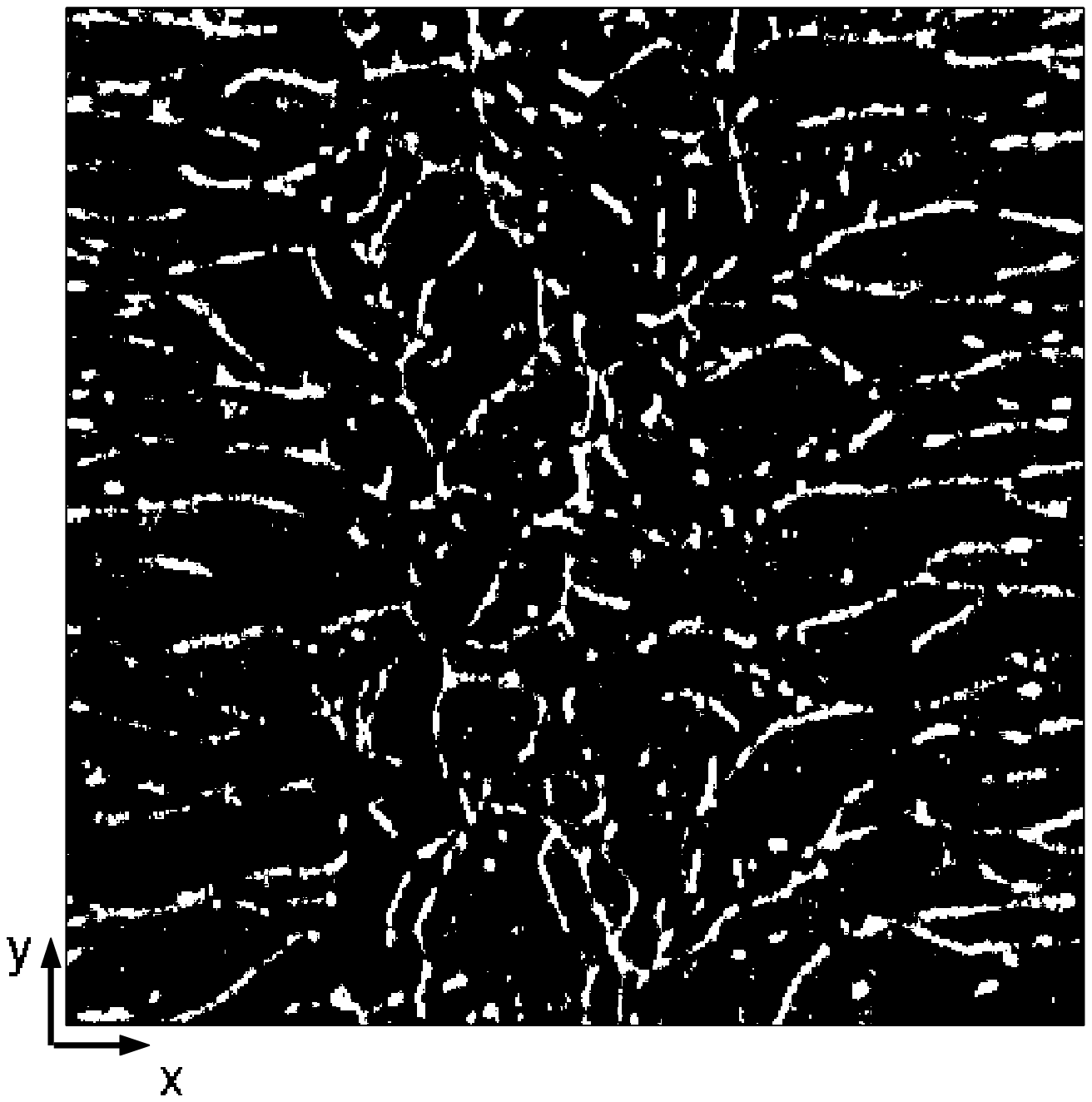}\label{fig:25j}}
  \hfill \subfigure[140s membrane in 15 cm$\times$15 cm tank,
  $Ra$ = 5$\times 10^{11}$, $\Delta C$ = 7.87 g/l, $Q$ = 0.39 mg/cm$^2$/min,
  $W_*$ = 0.47 cm/s, $Re$ = 1210.5, $V_I$ = 0.002 cm/s, Image area = 8.61 cm
  square, 528$^2$pix
  ]{\includegraphics[width=0.3\textwidth]{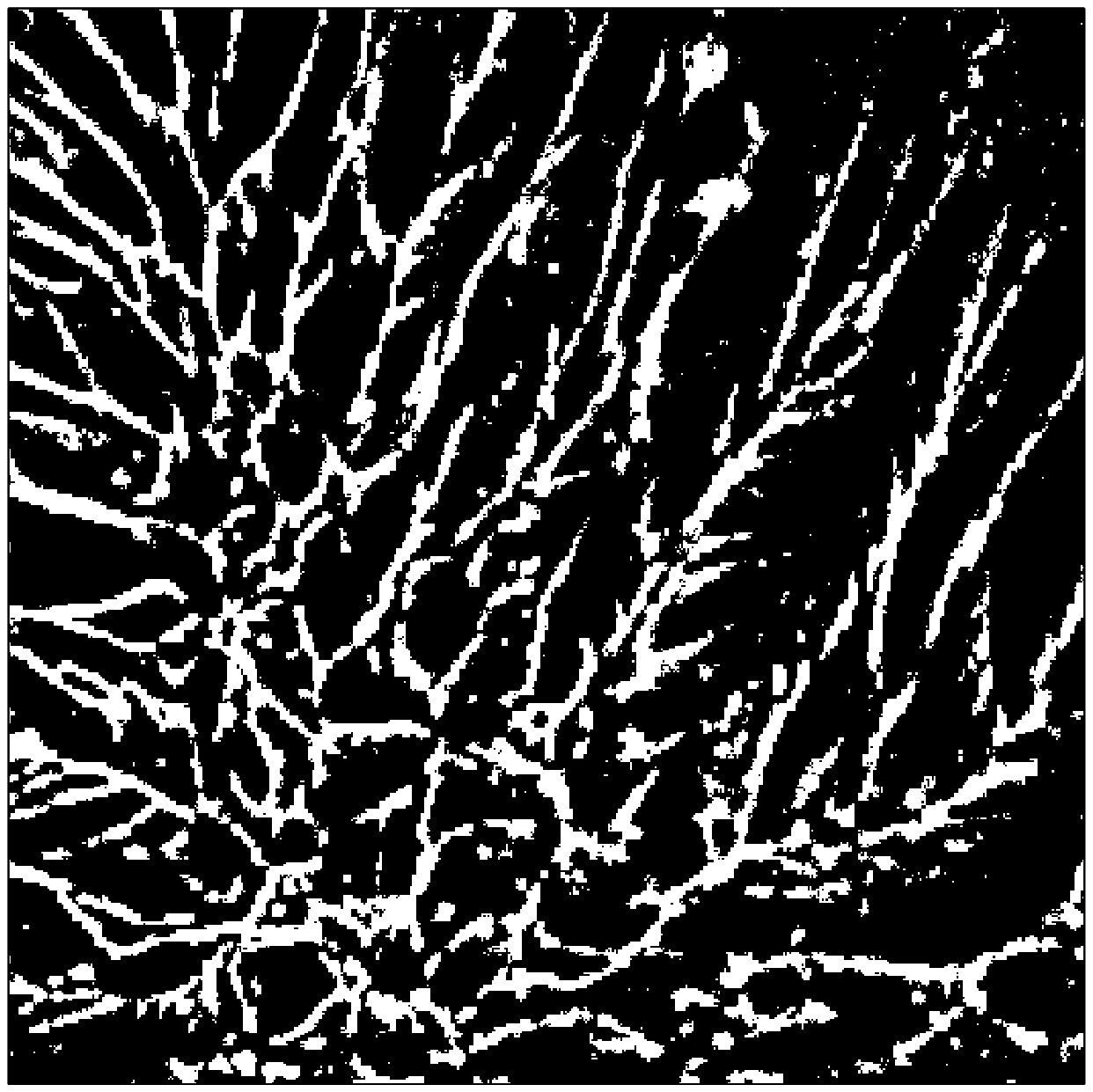}\label{fig:9m}}
  \hfill \subfigure[140s membrane in 15 cm$\times$15 cm tank,
  $Ra$ = 4.49$\times 10^{11}$, $\Delta C$ = 7.06 g/l, $Q$ = 0.41 mg/cm$^2$/min,
  $W_*$ = 0.48 cm/s, $Re$ = 1236.3, $V_I$ = 0.0023 cm/s, Image area = 8.32 cm
  square, 520$^2$ pix]{\includegraphics[width=0.3\textwidth]{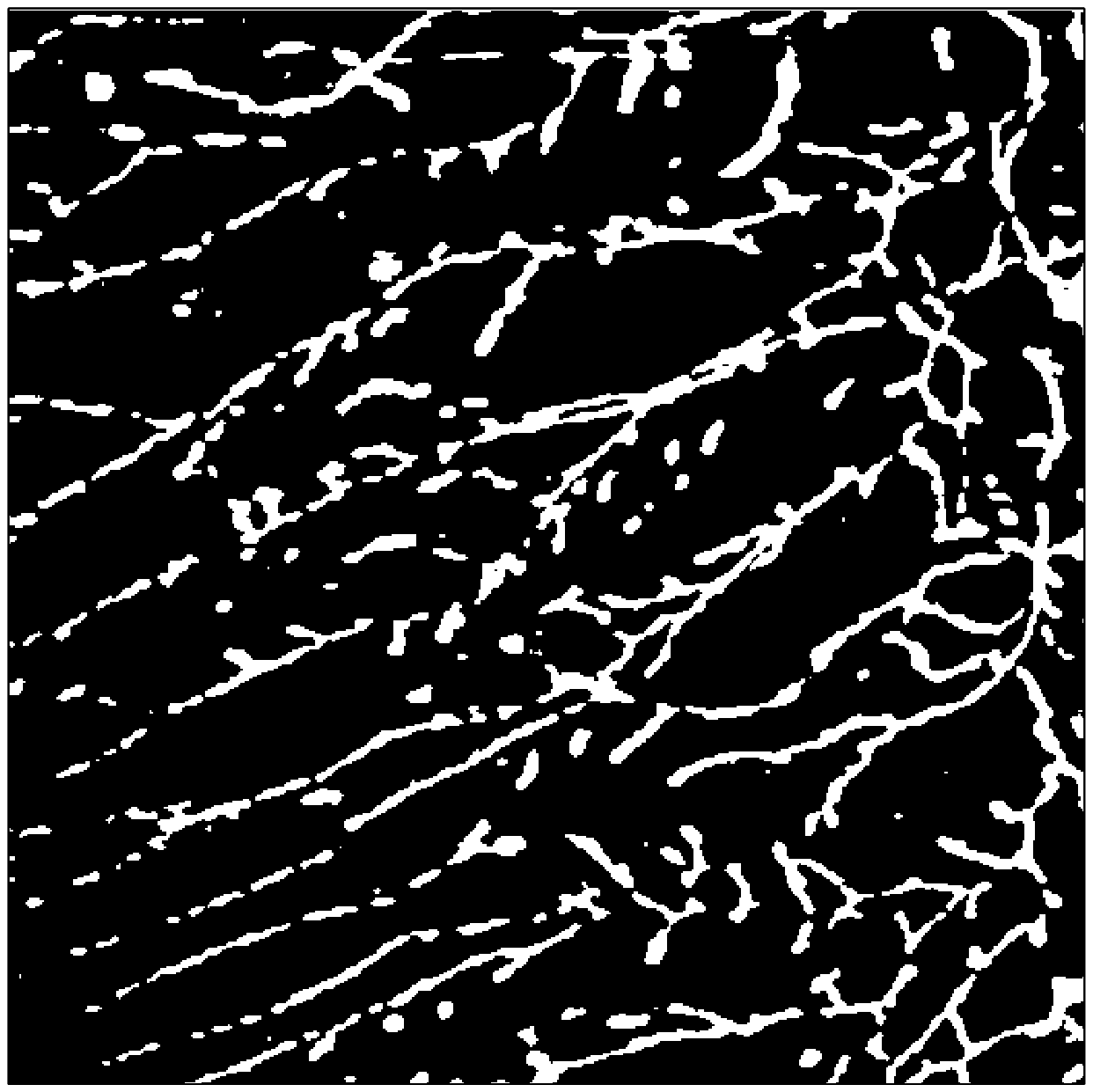}\label{fig:8jimg}}
  \subfigure[140s membrane in 15 cm $\times$15 cm tank, $Ra$ = 5.36$\times
  10^{11}$, $\Delta C$ = 8.42g/l, $Q$ = 0.41 mg/cm$^2$/min, $W_*$ = 0.48 cm/s,
  $Re$ = 1236.3, $V_I$ = 0.0016 cm/s, Image area = 7.58 cm square, 616$^2$ pix
  ]{\includegraphics[width=0.3\textwidth]{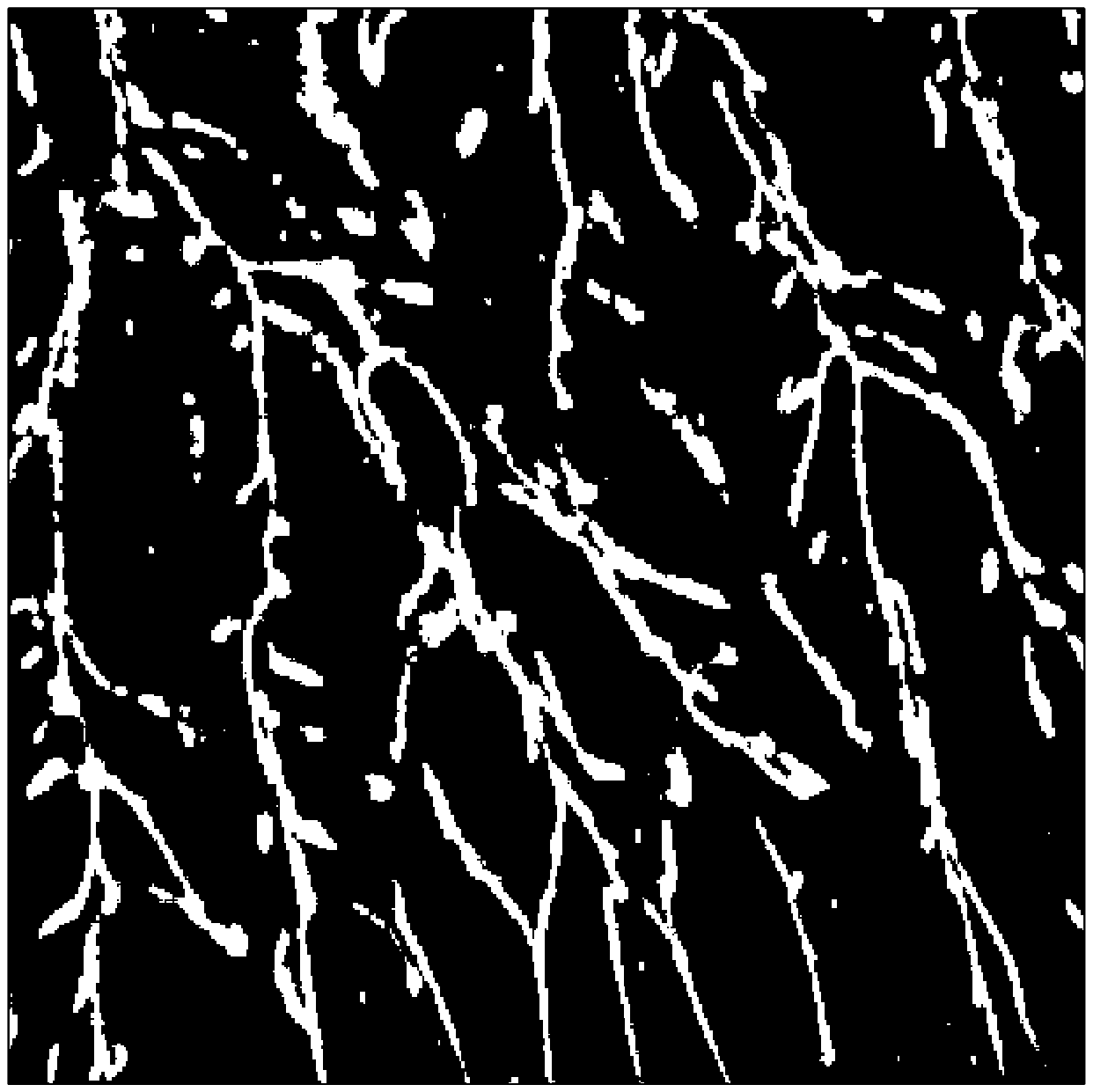}\label{fig:20m1}}\hspace{0.3cm}
  \subfigure[140s membrane in 15 cm$\times$15 cm tank, $Ra$ = 1.65$\times
  10^{11}$, $\Delta C$ = 2.6~g/l, $Q$ = 0.07 mg/cm$^2$/min, $W_*$ =
  0.27cm/s, $Re$ = 695.4, Image area= 14.69 cm square, 714$^2$ pix
  ]{\includegraphics[width=0.3\textwidth]{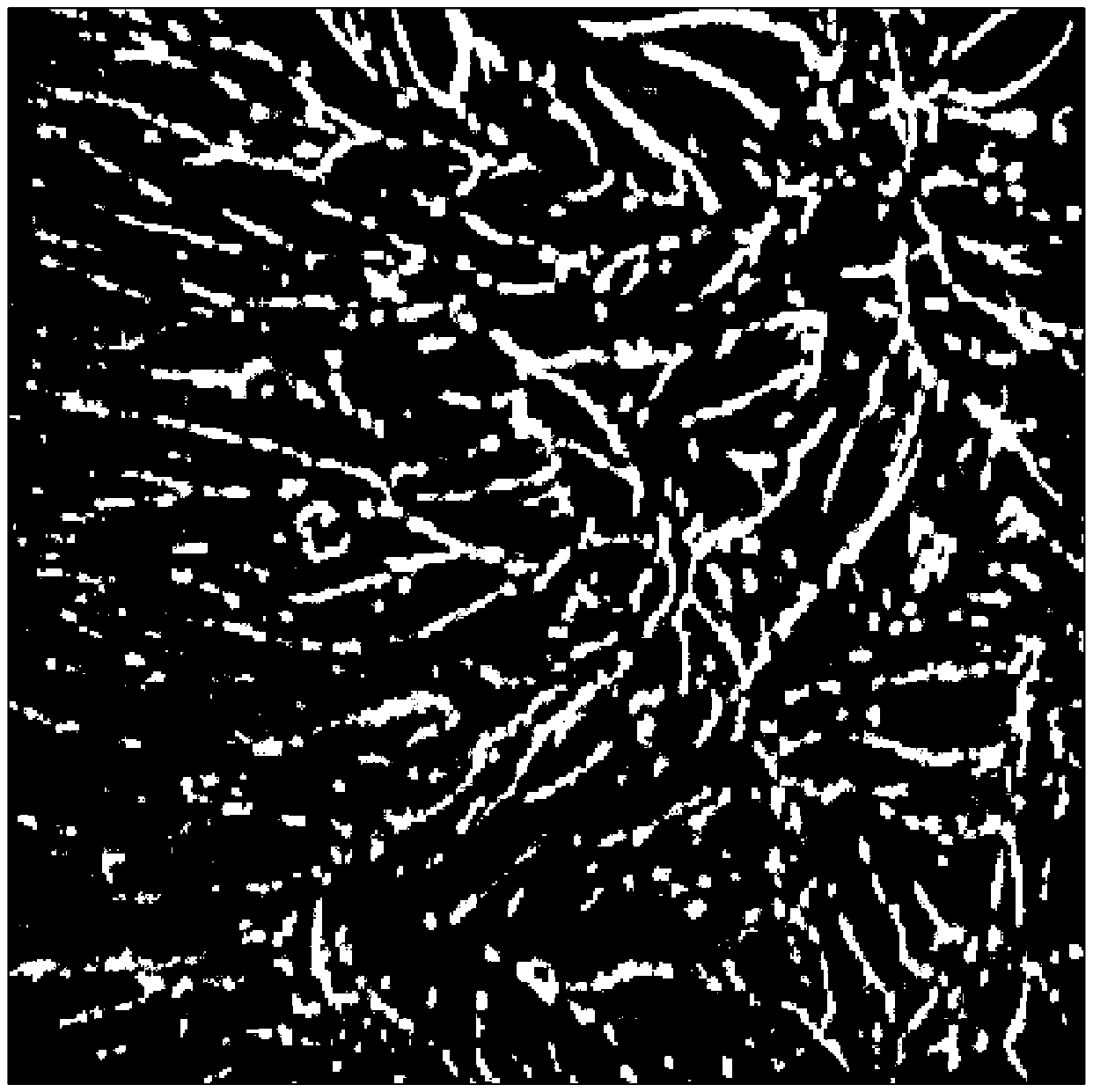 }\label{fig:13o}}
  \hfill
  \caption{The plan forms of near wall plume structures obtained under different conditions. (a)to (d) are for the D-Type showing full tank cross section. (e) to (g) show the planforms of the plume occupied region in the TF-type convection. (e) shows the corner area  with the walls only on the left and the bottom of the image. (f) shows the corner portion with the  walls on the top and the right. (g) shows a zoomed view of the central shear dominant portion. (h) shows the structure when the D-type occurs in the coarser membrane at lower driving potentials}
  \label{fig:planform}
 \end{figure}
 Figure~\ref{fig:planform} shows the binary images of the various
 planform plume structures obtained in the experiments. The parameter
 values are shown below each image.  $\Delta C $(g/l) is the effective
 driving concentration difference on the side of the membrane where
 the structure is visualised; $Ra$ is based on this $\Delta C $.  The
 white lines in the figures, of thickness $\sim 1$mm, are the bases of
 the sheet plumes originating from the membrane surface.  Two types of
 convection are observed depending on the pore size of the membrane.
 In one case, the transport of salt through the membrane is by
 diffusion (D-type henceforth). In the second case, the transport is
 due to a flow, albeit weak, across the membrane (TF type henceforth).
 In both cases, the line plumes emanate from the thin unstable
 boundary layers above the membrane.  As the focus of this paper is on
 the spatial scaling of these structures, we give only a brief
 description of the phenomena in each image of Figure
 \ref{fig:planform}; the details are discussed in \cite{mine}.
\subsection{Planforms in D-Type convection}
\label{sec:d-type} 
For the lower pore sized (PG $0.45\mu$) membrane, the transport across
the partition is diffusion dominated, while the transport above and
below the membrane is similar to Rayleigh - B\'{e}nard convection at
high $Ra$.  Figures~\ref{fig:10j1} to \ref{fig:25j} show the
structures obtained in the D-type convection. Similar plumes 
cover the  cross section on the lower side of the membrane.
The first two images are from the larger AR =0.65 which had multiple
large scale flow cells, the signatures of which are seen as circular
patches with aligned line plumes oriented radially around the patches,
the plume free circular patch being the area where the large scale
circulation impinges on the membrane. The diffusion layer thickness
$\delta = D\Delta C /Q$ is about 0.26 mm for the image of Figure 4(a).
The large scale flow cells shift their position at a larger time scale
($\sim$ 20 min) than one large scale circulation time ($\sim$ 1 min),
which is greater than the time scale of few seconds for the merging of
plumes.  Figure~\ref{fig:18d} shows the planform of plume structure
under the same conditions as in Figures ~\ref{fig:10j1} and
\ref{fig:10j2}, but in the smaller (10 cm $\times$ 10 cm, AR=0.435)
cross sectional area tank.  In this case, there was only a single
large scale circulation which was rotating clockwise in the $x-z$
plane and aligning the near wall plumes in the $x$-direction (See
Figure \ref{fig:25j} for the co-ordinate directions, $z$ is normal to
the membrane). The plume structure at lower $Ra$ in the larger AR
=0.65 tank, shown in Figure ~\ref{fig:25j}, had two counter rotating
circulation cells ( anticlockwise on the left and clockwise on the
right) rotating in the $x-z$ plane which created a near wall mean
shear directed toward the center along the $x$ direction. In all these
images, the flux scaled approximately as $\Delta C ^{4/3}$, equivalent
to $Nu\sim Ra^{1/3}$ in turbulent Rayleigh - B\'{e}nard convection.
\subsection{Planforms in TF-Type convection}
\label{sec:tf-type} 
Experiments with the coarser ($140s,\,35\mu$) membrane showed that
only half the area above the membrane was covered by plumes, with the
other half having plumes below the membrane.  This structure is due to
a weak through flow across the membrane.  The lower fluid dynamic
resistance of the coarser membrane allows this through flow. The
region having plumes on top of the membrane corresponds to upward
through flow and vice versa ( Figure \ref{fig:setup}(b)). The through
flow velocities $V_I$ are about 10 times smaller than the near wall
velocity scales in turbulent free convection given by $W_o\sim
\left(g\beta Q D\right)^{1/4}$\cite*[][]{ts1}.  Figure~\ref{fig:9m} to
\ref{fig:20m1} show the plan form view of the plume occupied region in
the TF type convection at AR =0.65.  Figure~\ref{fig:9m} shows the
corner region of this structure when there was a near wall mean shear
toward the bottom left corner due to a diagonally oriented large scale
circulation rotating in the clockwise direction.  Figure
\ref{fig:8jimg} shows a similar corner view of the structure when
there is a near wall mean shear along the diagonal toward the top
right corner, created by an anticlockwise circulation. The central
zoomed view, where the mean shear effects are predominant, in a
structure similar to that in Figures~\ref{fig:9m} and \ref{fig:8jimg},
is shown in Figure~\ref{fig:20m1}.  In all these cases, the flux
scaled as $\Delta C^3$ due to the presence of a flow across the
membrane. The phenomenology behind this flux scaling is described
in~\cite{mine}. At lower driving potentials, the TF-type convection in
the 140s membrane experiments was seen to change to the D-type with
$Q\sim \Delta C ^{4/3}$ scaling.  Figure~\ref{fig:13o} shows one such
type of plume structure.

In both the types of convection, the plume structures are continuously
evolving spatio-temporal patterns. But, as the time scale of the
merging of plumes is much smaller than the time scale of change of
Rayleigh number, the planforms could be expected to exhibit
statistically stationary characteristics for a given set of
parameters.  In addition, the scale free, non uniform and dendritic
appearance, along with the the common probability distribution of the
spacings, motivated us to analyse the structures using multifractal
analysis.
\section{Multifractal analysis}
\label{sec:mult-analys}
\subsection{Methodology and Scaling range}
\label{sec:meth-scal-range}
We use the standard box counting methodology to estimate the
multifractal exponents. The analysis is done on square binary images
(L$^2$ pixels). The measure $P_{i,j}$ where, $i,j$ are the box
indices, is the probability of occurrence of the plume in a box,
calculated as the ratio of white pixels in each box to the total
number of white pixels in the image. For each moment $q$, the
partition function $Z_q =\sum_{i,j}P_{i,j}^q$ is calculated for
various box sizes ($l_b$ in pixels). The slope of $\ln Z_q $ Vs
$\ln\left(\frac{l_b}{L}\right)$ gives the Cumulant generating function
$\tau(q)$ defined through $ \sum _{i,j}P_{i,j}^q\sim
\left(\frac{l_b}{L}\right)^{\tau(q)}$ where, $\tau(q) = (q-1) D_q$
with $D_q$ referring to $q^{th}$ order Renyi dimension.  The number of
boxes where $P_{i,j}\sim\left(\frac{l_b}{L}\right)^\alpha $ has a
singularity strength between $\alpha$ and $\alpha + d\alpha$ is given
by $ N(\alpha) \sim \left(\frac{l_b}{L}\right)^{-f(\alpha)}$.
$f(\alpha)$ is the fractal dimension of the subset - picked out by
each value of the moment $q$ - with singularity strength $\alpha$. We
calculate the H\"{o}lder exponent $\alpha$ and $f(\alpha)$ using the
direct method due to ~\cite{ac}. The expressions for $f(\alpha)$ and
$\alpha$ are $
  f(q)\,=\,\lim_{l_b\rightarrow 0}\frac{\sum_{i,j}\mu_{i,j}\,\ln\,\mu_{i,j}}{\ln(l_b/L)},\qquad
\alpha(q)\,=\,\lim_{l_b\rightarrow 0}\frac{\sum_{i,j}\mu_{i,j}\,\ln\,P_{i,j}}{\ln(l_b/L)},
$where $\mu_{i,j}\,=\,\frac{P_{i,j}^q}{\sum_{i,j}P_{i,j}^q}$ are the
normalized measures.  For each value of the moment $q$, the slope of
the of the plots of the numerator vs the denominator in the
expressions for $f(q)$ and $\alpha(q)$, in the range of values where
the plots are linear, give $f(q)$ and $\,\alpha(q)$. In our work, the
range of $q$ values is limited to $-2\le q\le 5$ in order to have a
reasonable range of scaling regime.
 
\begin{figure}
  \centering \subfigure[Range for $f(\alpha)$
  ]{\includegraphics[width=0.315\textwidth]{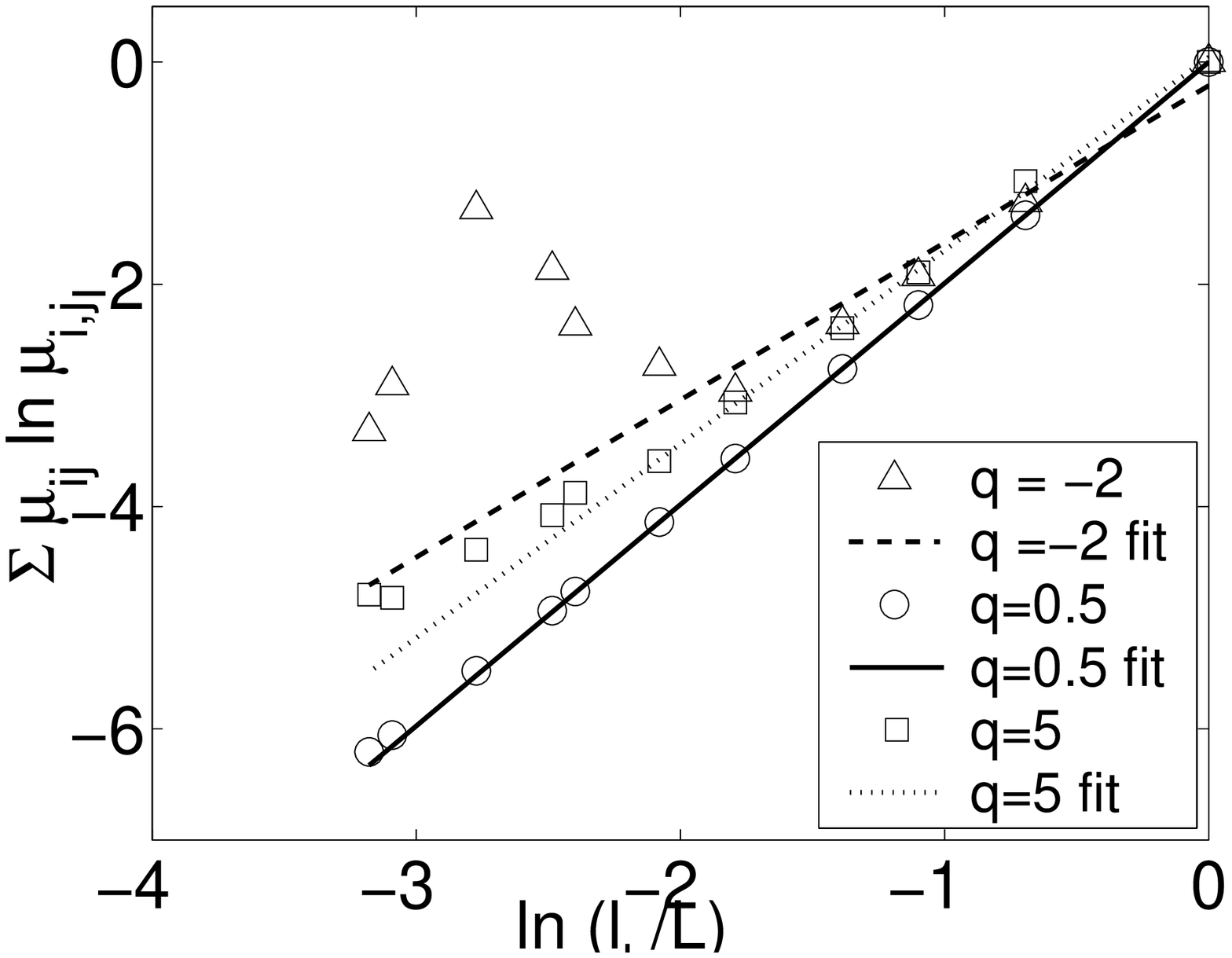
    }\label{fig:9mf}}\hfill \subfigure[Range for $\alpha$]{
    \includegraphics[width=0.315\textwidth]{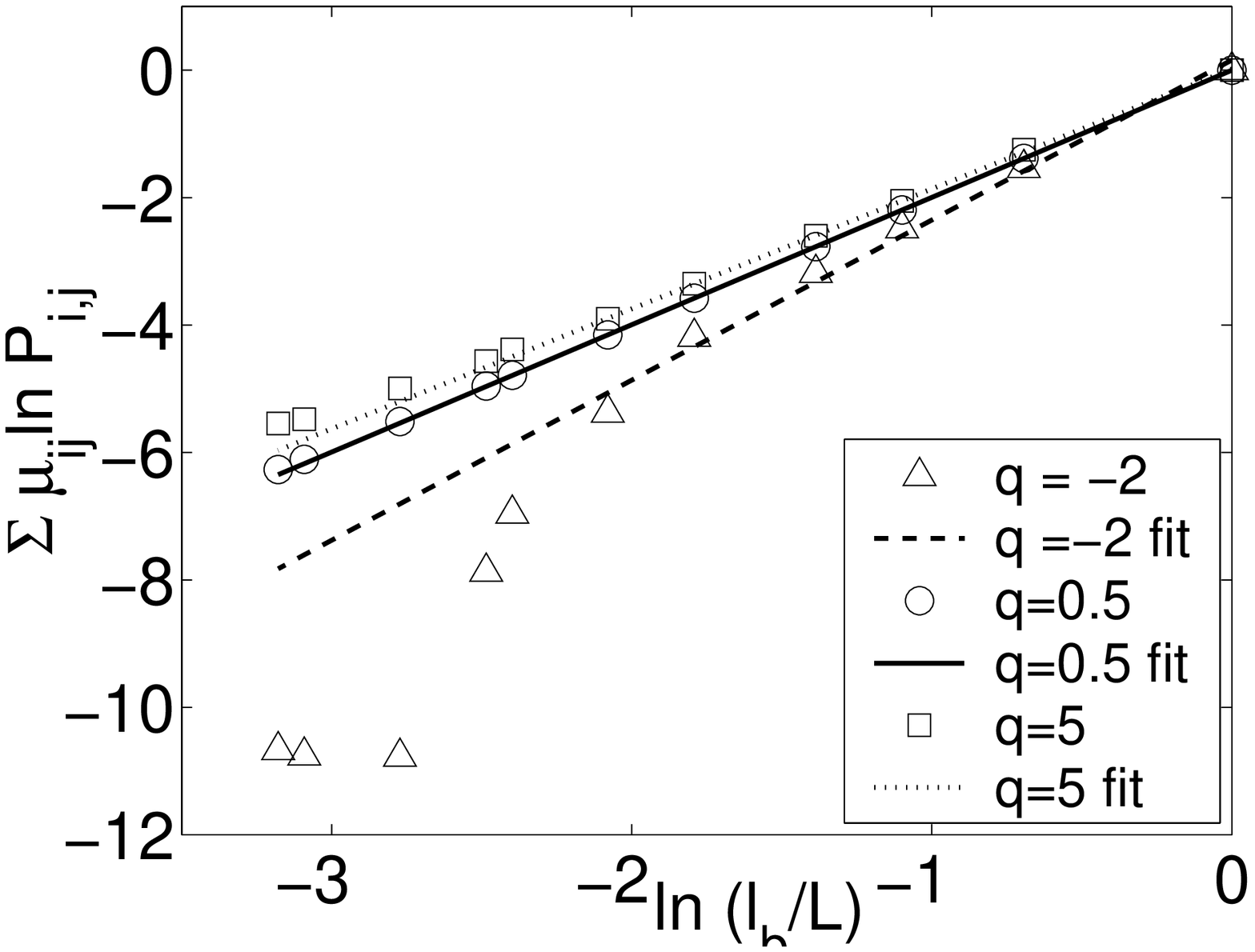
    }\label{fig:9mfs}}\hfill \subfigure[Range for $\tau(q)$]{
    \includegraphics[width=0.315\textwidth]{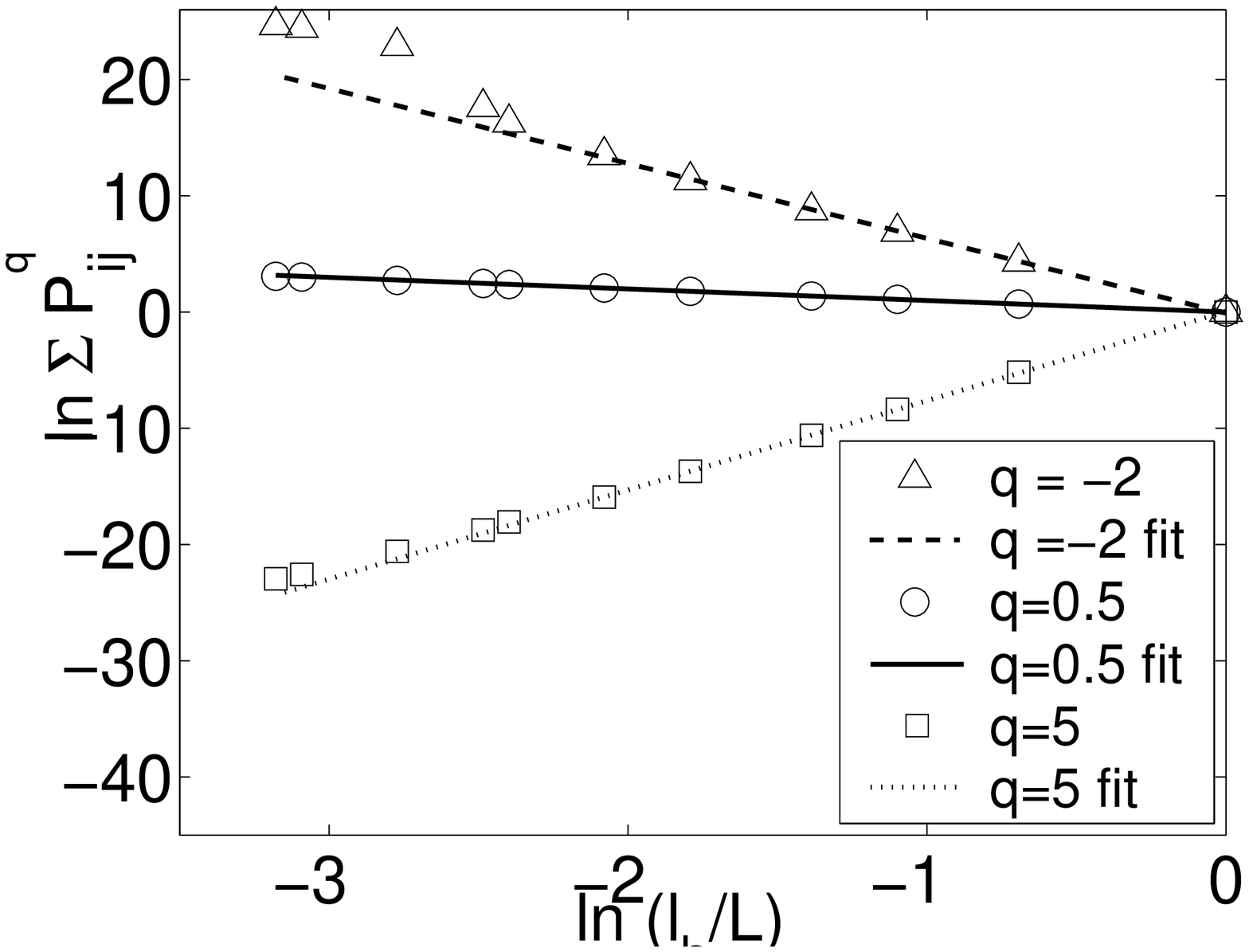}\label{fig:9mt}}\hfill
  \caption{The range of multifractal scaling for moments q = -2, 0.5 and 5. The lines show the linear fit used for calculating the slopes.} 
  \label{fig:9mslopes}
\end{figure}

The estimation of $f(\alpha)$ and $\alpha$ for the image in Figure
\ref{fig:9m} is shown in Figure \ref{fig:9mslopes}.  The image size is
8.61 cm square sampled at 528$^2$ pixels (L= 528). The calculation is
done for box sizes $l_b$ = 528, 264, 176, 132, 88, 66, 48, 44, 33, 24
and 22 pixels.  Figure~\ref{fig:9mf} shows the plots of
$\sum_{i,j}\mu_{i,j}\,\ln\,\mu_{i,j}$ versus $\ln(l_b/L)$ for a few
typical values of $q=-2,0.5$ and 5. The slope of the linear fit in the
range 66 pix $\le l_b\le$ 528 pix, (i.e., from 1.076 to 8.61 cm) is
used to estimate $f(\alpha(q))$ for all q. This slope is valid for a
larger length scale range of 22 pix $\le l_b\le$ 528 pix, (i.e.  from
0.359 to 8.61 cm ) for $q$ =-0.5 and 5.0, while for $q$ = -2, the
linearity holds only in the fit range.  The same linearity ranges for
various $q$ values are used for estimating $\alpha(q)$ from the plot
of $\sum_{i,j}\mu_{i,j}\,\ln\,P_{i,j}$ vs $\ln(l_b/L)$
(Figure~\ref{fig:9mfs}) and $\tau(q)$ from the plots of
$\ln\sum_{i,j}P_{i,j}^q $ vs $\ln (l_b/L)$ (Figure~\ref{fig:9mt}).
Figure~\ref{fig:9mf} to \ref{fig:9mt} show that the deviation from the
linearity is pronounced for $q =-2$ below a specific box size of
around 1cm ($\ln (l_b/L) =-2$), while it is almost negligible for
positive moments. This behavior is also seen in other studies like
\cite*{abc} and \cite{ms}.  As negative moments amplify low measures,
and the effect of noise in the low measure regions at smaller box
sizes is substantial, we expect this deviation to be due to the
amplification of the errors. For the same reason, the negative moment
slopes were also the most affected when the binary image threshold was
changed (See inset of Figure~\ref{fig:alphadiffn}).

The analysis is carried out for all the images in
Figure~\ref{fig:planform}.  The major common observations from the
plots of $\sum_{i,j}\mu_{i,j}\ln\mu_{i,j} $ and
$\sum_{i,j}\mu_{i,j}\ln P_{i,j} $ vs $\ln(l_b/L)$ for these images are
as follows. (a) For moments $q\ge -1$, the linearity of the above
quantities is seen to hold till about 0.5 cm.  As positive moments
pick out areas of higher measure, these moments represent the main
plume structure.  Thus, the main plume structure shows a multifractal
scaling in the range of tank cross section to 0.5 cm i.e., a range of
$2^5$. (b) When negative moments are less than -1, the plots of these
quantities show a deviation from linearity at about 1 cm in all the
images.  Hence, for $q < -1$, the multifractal scaling is valid for a
shorter length scale range of tank cross section to 1 cm, i.e.  $\sim
2^4$. (c) There seems to be a further lower cut off of approximately 5
mm, below which the slope of $\sum_{i,j}\mu_{i,j}\ln P_{i,j} $ Vs
$l_b/L$ for $q\le1$ goes to zero (See Figure \ref{fig:9mfs} at
$\ln(l_b/L)\sim -2.75$). This is expected to be because the box sizes
have reached the same order as the plume spacings. The geometric
structure of the planform does not exist below this length scale.
\begin{figure}
  \centering \includegraphics[width=0.6\textwidth]{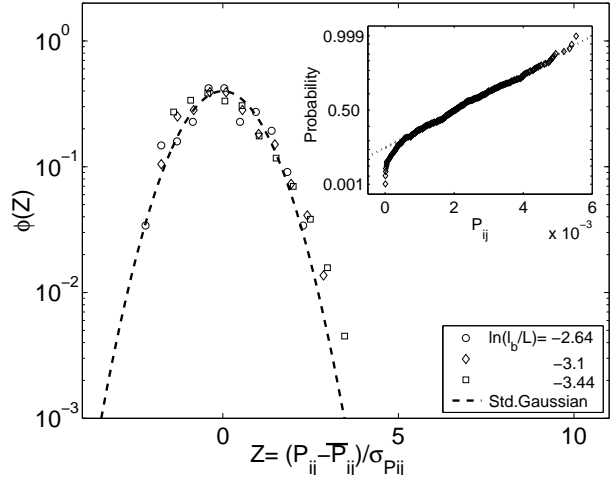}
\caption{The probability distribution function of the measure $P_{i,j}$ in the standardised form at three box sizes. The inset shows the normal probability plot of the measure at the intermediate box size.}
\label{fig:pdfmeas}
\end{figure}
To understand the above behaviour, we studied the probability
distribution function of the measures at various box sizes. The
distribution function of the measures is described by its moments
$Z_q$, and hence is related to $\tau(q)$ through $Z_q\sim
\left(l_b/L\right)^{\tau(q)}$. The distributions of $P_{ij}$ in the
standardised form $Z=\left(
  P_{ij}-\overline{P_{ij}}\right)/\sigma_{P_{ij}}$ for Figure
\ref{fig:13o}, where $\overline{P_{ij}}$ indicates the mean and
$\sigma_{P_{ij}}$ the standard deviation, for three box sizes are
shown in Figure~\ref{fig:pdfmeas}. The measure is distributed normally
for the larger box sizes. With decreasing box sizes, the low measure
and the high measure tails of the distribution deviates from the
Gaussian.  This is clear from the inset in Figure \ref{fig:pdfmeas},
which shows the normal probability plot of $P_{ij}$ for the
intermediate box size; the deviation from linearity shows the
difference from the Gaussian.  All the images in Figure
\ref{fig:planform} showed similar form and behaviour of the
probability distribution of the measures.  The deviation from
linearity for negative moments in Figure \ref{fig:9mslopes} at smaller
box sizes seems to be consistent with the deviation of the low measure
tails of the distribution function in Figure \ref{fig:pdfmeas}.
\subsection{The multifractal spectrum }
\label{sec:mult-spectr-}
\begin{figure}
  \centering 
\includegraphics[width=0.6\textwidth]{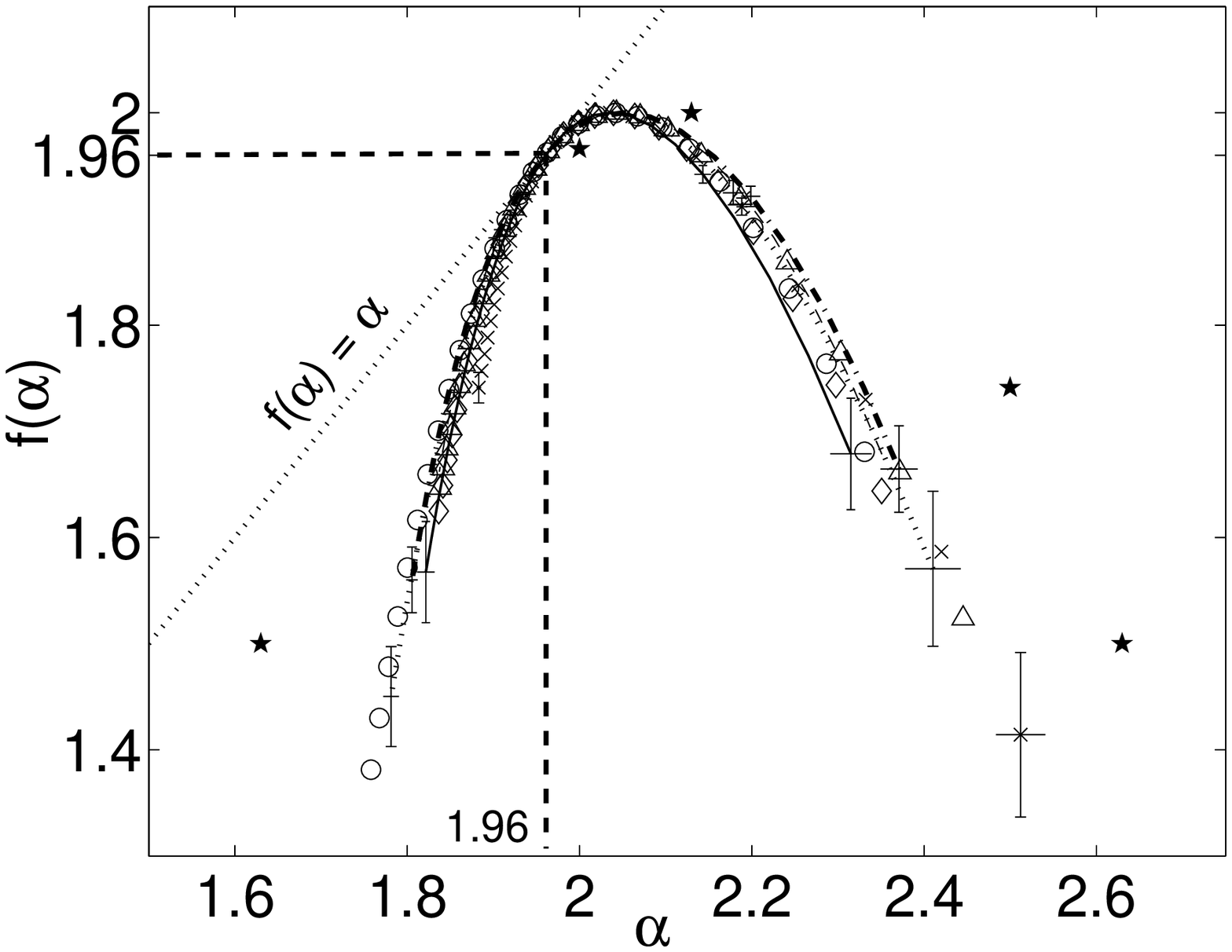}
  \caption{The $f-\alpha$ spectrum for all the
    images.  \textbf{....} Figure~\ref{fig:10j1}, \textbf{-.-.-}
    Figure~\ref{fig:10j2}, \textbf{$\circ$} Figure~\ref{fig:18d},
    \textbf{---} Figure~\ref{fig:25j}, \textbf{$\times$}
    Figure~\ref{fig:9m}, \textbf{$\lozenge$} Figure~\ref{fig:8jimg},
    $\bigtriangleup$ Figure~\ref{fig:20m1}, \textbf{- - -}
    Figure~\ref{fig:13o}, \textbf{$\star$} \cite{ms}.}
\label{fig:allfalpha} 
\end{figure}
\begin{table}
  \centering
  \begin{tabular}[c]{p{1cm}cccccccccccccccccc}
 $q$&  -2.0 &   -1.5&     -1.0&
      -0.5&    0&    
    0.5&        1.0&       1.5&
       2.0&     2.5&   
    3.0&      3.5&    4.0&
    4.5&    5.0\\
  $D_q$&  2.13&   2.09&      2.05&
      2.02 &       2.0&    
    1.98&      1.96&        1.95&
       1.93 &      1.92&   
    1.91&      1.897&        1.886&
       1.875 &     1.86
  \end{tabular}
  \caption{Variation of the Renyi dimension $D_q$ with the moment $q$ for Figure \ref{fig:10j1}}
\label{tab:dq}
\end{table}
The $f-\alpha$ curves of all the images in Figure~\ref{fig:planform}
are shown in Figure~\ref{fig:allfalpha}.  The error bars in the figure
are the errors in the slope of the linear curve fits.  The value of
$f$ at $q\, =\, 0$ is the fractal dimension of the `support' of the
measure. As the underlying area over which the plumes are formed is a
non-fractal surface, $f(\alpha(0))\,=\, 2$.  Figure
\ref{fig:allfalpha} shows that this is satisfied, suggesting a good
level of confidence in the calculation.  We also note that the fractal
dimension corresponding to the (information) entropy
$S\,=\,\sum_{\beta=0}^{b-1} p_\beta \log_b p_\beta,$ of the underlying
process which generates these structures is 1.96. The corresponding
values $D_q$ for Figure \ref{fig:10j1} are shown in
Table~\ref{tab:dq}. 

The results of our analysis of the near wall plume structure in
turbulent free-convection at high $Ra$ are strongly suggestive of its
multifractal nature. The analysed images cover a wide range of
conditions viz., $Ra$ range of about a decade, different test section
cross sectional areas (10$\times$10 cm$^2$ and 15$\times$15 cm$^2$),
absence and presence of a flow through the membrane, different large
scale flow strengths (about 2.5 times), a wide range of flux (over two
decades) and show a wide variety of structures having single and
multiple large scale flow cells with aligned and random structures.
The plot of Figure ~\ref{fig:allfalpha} show that, within the errors
encountered in the current analysis, all these images have the same
multifractal scaling of the main plume structure for length scales
greater than 0.5cm. 
\section{Discussion and Conclusions}
\label{sec:concl-disc}
\begin{figure}
\centering 
    \includegraphics[width=0.6\textwidth]{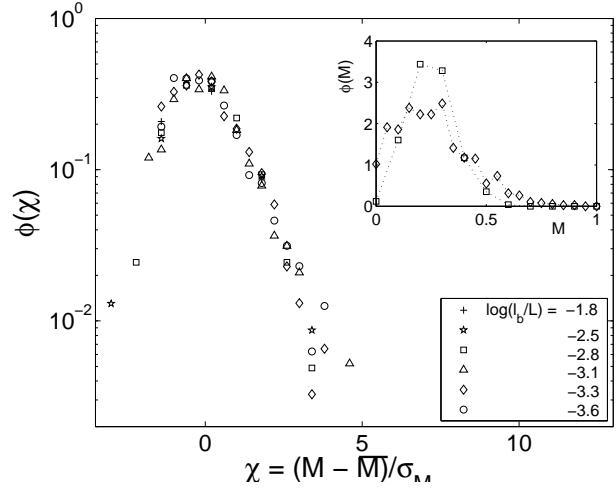}
\caption{The probability distribution function of the standardised form of the multipliers at different box sizes. The inset shows the distribution function of multipliers at two box sizes.}
\label{fig:multpdf}
\end{figure}

We have shown that the plan forms of near wall plume structure in high
$Ra$ turbulent free-convection under varying parameter values have the
same multifractal spectrum of singularities. Even though the plan
forms appear substantially different in their geometric form, they
have the same non trivial spatial scaling. The probabilistic
interpretation of multifractal formalism, as described by \cite{mb2},
shows that $f$ and $\alpha$ indirectly describe the underlying
generating mechanism of these structures in terms of the multipliers
$M$ for each stage of the process. These exponents are independent of
the length scales and are unique functions of the multiplier
distribution which decides how the measure is distributed at each
stage.  Figure~\ref{fig:multpdf} shows the probability distribution of
the multipliers in the standardised form, $\chi=
\left(M-\overline{M}\right)/\sigma_M$.  The multiplier values are
calculated as the ratio of the measures in each box to the measure in
its parent box, when each box divides into four boxes. The
calculations are done for a series of box sizes as shown in the legend
of the Figure~\ref{fig:multpdf}.  The figure shows that the image has
the same form of distribution function of multipliers at different
length scales, which has an exponential tail of higher multipliers.
Note that the length scale spans the entire range over which
$f(\alpha)$ has been obtained. Similar results were obtained for all
the images in Figure \ref{fig:planform}. In physical terms, this means
that the present images have the same underlying generating mechanism.
The common generating mechanism that can be identified here is that
the thin boundary layers of lighter fluid above the membrane becomes
unstable resulting in the generation of sheet plumes.  Interaction of
the neighboring plumes due to entrainment leads to their merger. New
plumes are nucleated in the vacant space due to the merger. The whole
process is also influenced by the external flow field created by the
mean wind, whose effect is to align the plume lines in the direction
of the mean shear.

An identical probability distribution of multipliers at different
stages of the process would show that the multipliers at different
stages are not correlated and the underlying process to be
statistically self similar ~\cite*[][]{mb2,ackrs,rdf}. The inset in
Figure \ref{fig:multpdf} shows the probability distribution function
of the multipliers, shown only at two box sizes for clarity. The
distribution functions coincide at larger multiplier values for the
whole multifractal range, but are different at low multiplier values.
We expect this to be a due to the correlated nature of the ends of the
structure (low multipliers), which move towards the main plume lines
due to entrainment.
\begin{figure}
 \centering
  \includegraphics[width=0.6\textwidth]{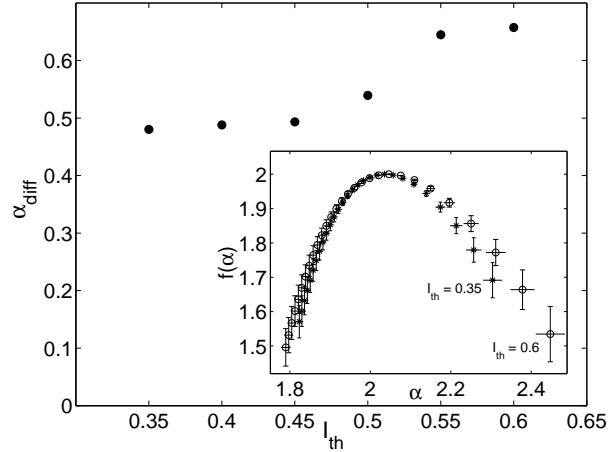}
  \caption{The variation of $\alpha_{diff}= \alpha_{max}-\alpha_{min}$
    with the threshold $I_{th}$ for Figure \ref{fig:25j}. The inset
    shows the $f-\alpha$ curve for the two extreme thresholds}
\label{fig:alphadiffn}
\end{figure}

Recall that for multinomial processes, $f(\alpha)$ is calculated as
the maximum from a range of values for any $\alpha$ for each moment
$q$.  Connecting this with the thermodynamic analogy noted in the
literature \cite[][]{abc}, the common $f-\alpha$ curves might
imply that the plume structure in turbulent convection is formed so as
to maximize the entropy of the structure.

Earlier studies on the multifractal nature of energy dissipation field
in turbulent flows by \cite{ms} have parallels with the current study.
The energy dissipation is caused by the velocity gradients at the
viscous scales. Plume edges represent the major gradients of
velocities near the wall in convection. These also are the relevant
viscous scales near the wall. Hence, it is possible that the near wall
energy dissipation field follows the plume structure closely. The
$f-\alpha$ curves obtained by \cite{ms} are shown in comparison with
the present curves in Figure \ref{fig:allfalpha}. The $f(\alpha)$
curves obtained in our case have a lower spread.  Further work is
needed to clarify the possible connection of these energy dissipation
studies to the present analysis.
\appendix
\section{Effect of thresholding}
\label{sec:effect-thresholding}
The robustness of the calculated multifractal exponents depends on how
representative the binary image is of the RGB LIF image.
Figure~\ref{fig:alphadiffn} shows the variation of $\alpha_{diff}=
\alpha_{max}-\alpha_{min}$ with the threshold $(0 \le I_{th}\le 1)$
for the image in Figure~\ref{fig:25j}. The inset in Figure
\ref{fig:alphadiffn} shows the $f-\alpha$ curves at the two extreme
thresholds. $I_{th}\ge 0.55$ produced binary images clearly missing in
finer details, while $I_{th}\le 0.45$ created images that were
noisier, with larger plume thickness.  Therefore, for this image, the
threshold was chosen to be 0.45. The variation of $\alpha_{diff}$ for
$0.45\ge I_{th}\le 0.55$ is about 0.15, of the same order as the error
in $\alpha$. Therefore, the change in the $\alpha$ due to the error
from the optimum threshold is within the error involved in the
calculation of $\alpha$.  Further, the inset of
Figure~\ref{fig:alphadiffn} shows that the threshold does not affect
the form of the $f-\alpha$ curve.  Hence in all the images, the
threshold was chosen by visual judgment along with sensitivity
analysis. The above analysis was for one of the low quality images;
for better quality images as in Figure~\ref{fig:8jimg}, there was
negligible variation with threshold.

We thank G.Vishwanath and Joby Joseph for assistance related to
image processing and K.R.Sreenivasan for his comments on the draft
version.  
\bibliography{babu}

\begin{thebibliography}{19}
\expandafter\ifx\csname natexlab\endcsname\relax\def\natexlab#1{#1}\fi

\bibitem[Bershadskii {\em et~al.\/}(2004)Bershadskii, Niemela, Praskovsky \&
  Sreenivasan]{berd}
{\sc Bershadskii, A., Niemela, J., Praskovsky, A. \& Sreenivasan, K.} 2004
  ``{C}lusterisation'' and intermittency of temperature fluctuations in
  turbulent convection. {\em Physical Review E\/} {\bf 69}, 056314.

\bibitem[Chabra {\em et~al.\/}(1989)Chabra, Meneveau, Jensen \&
  Sreenivasan]{abc}
{\sc Chabra, A., Meneveau, C., Jensen, R. \& Sreenivasan, K.~R.} 1989 Direct
  determination of the $f(\alpha$) singularity spectrum and it application to
  fully developed turbulence. {\em Physical Review A\/} {\bf 40}~(9),
  5284--5294.

\bibitem[Chabra \& Jensen(1989)]{ac}
{\sc Chabra, A.~B. \& Jensen, R.~V.} 1989 Direct determination of $f(\alpha$)
  singularity spectrum. {\em Physical Review Letters\/} {\bf 62}~(12),
  1327--1330.

\bibitem[Chabra \& Sreenivasan(1992)]{ackrs}
{\sc Chabra, A.~B. \& Sreenivasan, K.~R.} 1992 Scale invariant multipliers in
  turbulence. {\em Physical Review Letters\/} {\bf 68}~(18), 2762.

\bibitem[Deardorff(1970)]{dear1}
{\sc Deardorff, J.} 1970 Convective velocity and temperature scales for the
  unstable planetary boundary layer and for {R}ayleigh convection. {\em Journal
  of the Atmospheric Sciences\/} {\bf 27}, 1211--1213.

\bibitem[Frederiksen {\em et~al.\/}(1997)Frederiksen, Dahm \& Dowling]{rdf}
{\sc Frederiksen, R.~D., Dahm, W. \& Dowling, D.} 1997 Experimental assessment
  of fractal scale similarity in turbulent flows. part 3. multifractal scaling.
  {\em Jl. Fluid. Mech.\/} {\bf 338}, 127--155.

\bibitem[Halsey {\em et~al.\/}(1986)Halsey, Jensen, Kadanoff, Procaccia \&
  Shraiman]{hals}
{\sc Halsey, T., Jensen, M., Kadanoff, L., Procaccia, I. \& Shraiman, B.} 1986
  Fractal measures and their singularities:the characterisation of strange
  sets. {\em Physical Review A\/} {\bf 33}~(2), 1141--1151.

\bibitem[Kerr \& Herring(2000)]{kerr2}
{\sc Kerr, R.~M. \& Herring, J.~R.} 2000 {P}randtl number dependence of nusselt
  number in direct numerical simulations. {\em Jl. Fluid. Mech.\/} {\bf 419},
  325--344.

\bibitem[Mandelbrot(1989)]{mb2}
{\sc Mandelbrot, B.~B.} 1989 Multifractal measures, especially for the
  geophysicist. {\em Pure and applied geophysics\/} {\bf 131}~(1/2), 5--42.

\bibitem[Meneveau \& Sreenivasan(1991)]{ms}
{\sc Meneveau, C. \& Sreenivasan, K.~R.} 1991 The multifractal nature of
  turbulent energy dissipation. {\em Jl. Fluid. Mech.\/} {\bf 224}, 429--484.

\bibitem[Niemela {\em et~al.\/}(2000)Niemela, Skrbek, Sreenivasan \&
  Donelly]{niemn}
{\sc Niemela, J., Skrbek, L., Sreenivasan, K. \& Donelly, R.} 2000 Turbulent
  convection at very high {R}ayleigh numbers. {\em Nature\/} {\bf 404},
  837--840.

\bibitem[Puthenveettil(2004)]{mine}
{\sc Puthenveettil, B.~A.} 2004 Investigations on high {R}ayleigh number
  turbulent free-convetion. PhD thesis, Indian Institute of Science, Bangalore,
  http://www.mecheng.iisc.ernet.in/\textasciitilde apbabu/resinfo.htm.

\bibitem[Siggia(1994)]{siggr}
{\sc Siggia, E.~D.} 1994 High {R}ayleigh number convection. In {\em Ann Rev
  Fluid Mechanics\/}, , vol.~26, pp. 137--168.

\bibitem[Spangenberg \& Rowland(1961)]{sprow}
{\sc Spangenberg, W.~G. \& Rowland, W.~G.} 1961 Convective circulation in water
  induced by evaporative cooling. {\em Phys. Fluids\/} {\bf 4}~(6), 743--750.

\bibitem[Tamai \& Asaeda(1984)]{tas}
{\sc Tamai, N. \& Asaeda, T.} 1984 Sheet like plumes near a heated bottom plate
  at large {R}ayleigh number. {\em Jl. Geophys. Res.\/} {\bf 89}, 727--734.

\bibitem[Theerthan \& Arakeri(1998)]{tjfm}
{\sc Theerthan, S.~A. \& Arakeri, J.~H.} 1998 A model for near wall dynamics in
  turbulent {R}ayleigh - {B}\'enard convection. {\em Jl. Fluid. Mech.\/} {\bf
  373}, 221 --254.

\bibitem[Theerthan \& Arakeri(2000)]{tapf}
{\sc Theerthan, S.~A. \& Arakeri, J.~H.} 2000 Plan form structure and heat
  transfer in turbulent free convection over horizontal surfaces. {\em Phys.
  Fluids\/} {\bf 12}, 884--894.

\bibitem[Townsend(1959)]{ts1}
{\sc Townsend, A.} 1959 Temperature fluctuations over a heated horizontal
  surface. {\em Jl. Fluid. Mech.\/} {\bf 5}, 209--211.

\bibitem[Zocchi {\em et~al.\/}(1990)Zocchi, Moses \& Libchaber]{zocchi}
{\sc Zocchi, G., Moses, E. \& Libchaber, A.} 1990 Coherent structures in
  turbulent convection, an experimental study. {\em Physica A\/} {\bf 166},
  387--407.

\end{thebibliography}
\end{document}